# Will Einstein Have the Last Word on Gravity?

## Bernard F. Schutz[1], Joan Centrella[2], Curt Cutler[3], Scott A. Hughes[4]

## Abstract

This is a whitepaper submitted to the 2010 Astronomy Decadal Review process, addressing the potential tests of gravity theory that could be made by observations of gravitational waves in the milliHertz frequency band by the proposed ESA-NASA gravitational wave observatory LISA. A key issue is that observations in this band of binary systems consisting of black holes offer very clean tests with high signal-to-noise ratios. Gravitational waves would probe nonlinear gravity and could reveal small corrections, such as extra long-range fields that arise in unified theories, deviations of the metric around massive black holes from the Kerr solution, massive gravitons, chiral effects, and effects of extra dimensions. The availability of strong signals from massive black hole binaries as well as complex signals from extreme mass-ratio binaries is unique to the milliHertz waveband and makes LISA a particularly sensitive probe of the validity of general relativity.

15 February 2009

[1] Albert Einstein Institute, Potsdam, Germany (Bernard.Schutz@aei.mpg.de)
[2] NASA Goddard Space Flight Center
[3] Jet Propulsion Laboratory, California Institute of Technology
[4] Massachussets Institute of Technology

# Will Einstein Have the Last Word on Gravity?

## Gravitational wave observations probe strong-field nonlinear gravity

General relativity (GR) is a theory of gravity, one of the fundamental forces of nature, in which gravitational fields are manifested as curvature of spacetime. It is separate from the rest of theoretical physics in two ways: first, it is not a quantum theory, and second, it is described by a geometry that influences everything rather than by a force that acts selectively. Most attempts at incorporating gravity into a quantum framework also introduce extra gravity-like force fields, which act more selectively and provide evidence of a breakdown in general relativity even in situations far from the purely quantum scales of the Planck length or the Planck energy. Indeed, a central question in fundamental physics today is, *how does quantum gravity manifest itself at low energies and long distance scales?*

> **Fundamental tests must be clean tests**
> • The perihelion precession of Mercury opened the door to GR. The Newtonian orbit model, with planetary perturbations, had no free parameters.
> • The Hulse-Taylor Binary Pulsar is a simple two-body system whose orbit can be computed to high accuracy within GR. The GW energy loss has no free parameters.
> • Massive black hole binary systems have simple two-body orbits that are as well understood as the Binary Pulsar and which probe ultra-strong gravity as the components spiral together toward merger.
> • Stellar black holes captured by massive black holes have complex orbits but are still pure two-body systems. All parameters can be measured from the GW signal.

Gravitational wave (GW) observations, particularly in the milliHertz band that LISA will explore, offer the means to challenge GR at levels of precision and gravitational field strengths unthinkable even twenty years ago. Binary black holes are among the *cleanest* possible astronomical systems we can observe, and as such are ideally suited for making fundamental tests. The staggering simplicity of the Kerr metric, where an equilibrium macroscopic object is *exactly* described by just two parameters, its mass and spin, allows little freedom to fit or model away anomalous observations. Binary black holes are pure vacuum gravity, and their general relativistic orbits and even their mergers can be solved now with very high accuracy. No modeling of fluids, magnetic fields, or radiative transport is involved. Either the observations will agree with GR or the theory will fail the test.

> **Key science questions**
> • What can gravitational wave astronomy tell us about new physics?
> • How does quantum gravity manifest itself far below the Planck energy?
> • Are the massive dark central objects in galaxies really Kerr black holes?
> • Can naked singularities form?

The milliHertz GW band, where LISA will observe, allows such tests with unprecedented precision. In the hundreds of expected events during LISA's lifetime, the opportunity to see deviations from GR is very real. The black holes that radiate at these frequencies are massive, their signals strong, and LISA's signal-to-noise ratio (SNR) remarkably high, up to $10^8$ in energy[1]. The best tests will, most probably, be done with the signals that have the highest SNRs.

GR has passed all tests so far with no hint of a problem, and its success in explaining the Hulse-Taylor (H-T) binary pulsar assures us that in the LISA frequency band, only a decade above the H-T radiation, our source estimates and detector techniques are reliable. But the H-T system has orbital gravitational fields only slightly stronger than those in our solar system, so when LISA observes *nonlinear* gravity in tight black hole binaries, deviations from GR could show up for the first time.

It is worth reminding ourselves why and where GR might fail. Quantum gravity, as in string theory, might introduce new long-range fields that could significantly affect orbits and the emitted radiation in strong fields, or could introduce extra polarization states into the observed gravitational waves. Inflation itself and dark energy today could be evidence of failures in GR. (see the white paper "Gravitational Waves from New Forms of Energy"). Grand unified theories could well have an uncharged "shadow matter" sectors containing more than one cosmological dark matter particle, forming compact structures radiating GWs. Some speculative brane-world scenarios have similar effects. Signals traveling from high redshifts can accumulate tiny effects into measureable phase delays: extra chiral terms in the gravity Lagrangian could cause phase shifts between the two circular polarizations, or a tiny graviton mass could induce frequency dispersion. Even classical GR could yet break down: we still have no theorem excluding naked singularities, and if a Kerr black hole could manage to accrete too much angular momentum its horizon would disappear and expose its singularity.

To identify any deviation from GR requires a clean signal and a good SNR. *Any such failure of GR should point the way to new physics.*

> **Where will Einstein's theory fail?**
> At the dawn of the 20$^{th}$ century, Newton's theory of gravity was astonishingly successful. But precision measurements of the precession of the perihelion of Mercury's orbit about the Sun revealed a small discrepancy of 43 seconds of arc per century. Why Mercury? Because the planet probed the strongest gravitational fields accessible to 19$^{th}$ century astronomers. General relativity's first triumph was to account naturally for this anomaly.
>
> At the dawn of the 21$^{st}$ century, Einstein's theory of gravity is astonishingly successful. GR successfully predicts phenomena such as gravitational lensing, black holes, and gravitational waves, and provides a natural framework for the expansion of the Universe. But there is no well established quantum theory of gravity, and gravity is not yet unified with the other fundamental fields. We know GR must fail but we don't know how.
>
> A sensible strategy is to look where we have not been able to look before. We must make precision measurements where gravity is strong, highly nonlinear, and fully dynamical: black hole mergers. LISA's low-frequency gravitational wave observations provide ideal access to such gravitational fields.

---

[1] GW signal-to-noise ratios (SNR) are normally quoted as amplitude ratios, because the waves are emitted and detected coherently. But the energy SNR, which is the square of this, is more directly comparable, in terms of information content, to the SNR of normal optical astronomy, where the energy of photons and not their phase is detected. We will use energy SNRs in this paper.

## Experimental limits on violations of GR

While so far GR has passed all the tests to which we have subjected it (Will 2006), most of these tests have been in the weak-field regime, which we can define by using the parameter $v^2/c^2 \sim GM/(Rc^2)$ with $v$, $M$, and $R$ the bodies' typical velocity, mass, and separation. For many applications, such as Solar System dynamics, it is perfectly adequate to describe the dynamics using post-Newtonian (PN) equations, which are a weak-field slow-motion expansion of the full general relativistic equations in powers of $(v/c)^2$. For the tests of GR in our Solar System, it has so far been sufficient to use the first-order PN equations ($v^2/c^2 \sim 10^{-8}$) to get agreement with GR. Future tests may begin to probe second-order PN corrections.

Binary pulsars, which are essentially very stable and accurate clocks with typical orbital velocities $v/c \sim 10^{-3}$, have been excellent laboratories for precision tests of GR (Lorimer 2008). Current observations of several binary pulsars are consistent with GR using equations of the first PN order, plus GW radiation reaction, which is at 2.5 PN order ($v^5/c^5$). Observations of the first binary pulsar to be discovered (earning Hulse and Taylor the Nobel Prize), PSR B1913+16, provided the first clean astronomical test of gravitational radiation. Loss of GW energy (radiation reaction) causes the binary's orbit to shrink slowly; its period derivative $dP/dt$ agrees with that predicted by GR to 0.2%, well within the error bars (Weisberg & Taylor 2004).[2] The *double pulsar system*, PSR J0737-3039AB, will soon do even better (Kramer *et al.* 2006).

Note that the orbital gravitational fields in known binary pulsars are not much stronger than those in the solar system: the semimajor axis of the orbit of 1913+16 is about 1.4 $R_\odot$. These weak orbital fields limit their ability to probe nonlinear GR *dynamics*. They do provide important tests of strong-field *static* gravity—basically because the redshift at the surface of a neutron star is of order 0.2. While this strong internal gravity does not affect the orbital motion in GR (the strong equivalence principle), in some scalar-tensor theories there can be order unity departures from GR's predictions for the orbital motions (Damour 2000).

## The inspiral, merger, and ringdown of MBH binaries

LISA's strongest sources are expected to be coalescing massive black hole (MBH) binaries, where the components have roughly comparable masses, $0.1 < M_2/M_1 < 1$. The waveforms will be visible by eye in the data stream, standing up well above the noise even for high-redshift sources, as illustrated in Figure 1. The inspiral stage is a relatively slow, adiabatic process in which the MBHs spiral together on quasi-circular orbits. This part of the waveform can be computed analytically, with high-order PN expansions (order 3 or more). The inspiral is followed by the dynamical merger, in which the MBHs leave their quasi-circular orbits and plunge together, forming a highly distorted remnant MBH. Here, the velocities approach $v/c \sim 1/3$, the PN approximation breaks down, and the system can only be analyzed using numerical simulations of the full Einstein equations. The distorted remnant settles into a stationary Kerr black hole as it "rings down" by emitting gravitational radiation, which can be calculated analytically from BH perturbation theory. See Figure 2 for the range of LISA SNRs expected.

---

[2] Why is radiation reaction – an *O(2.5)* PN effect – needed while the *O*(2) PN effects are left out of the orbit model? Because it is a secular effect that accumulates with time: radiation reaction drains energy from the system and so causes inspiral, which makes the orbital phase grow quadratically with time. By contrast, even-order PN equations are conservative, and so their effects on the orbit are quasi-periodic, although their period (as for precession) can be so long that the effect at first grows linearly in time.

Each phase of the signal provides different information. During the inspiral phase, fitting the signal to the PN waveform determines the holes' masses (typically to better than 0.1%) and spins (to 1%) (Lang & Hughes 2006), plus orbital parameters like the inclination and eccentricity (if any: it is expected to be small). There will be plenty of time to predict the time, phase, and location of the merger and alert other astronomers who may want to do simultaneous observations. The merger waveform can have SNR comparable to the inspiral phase, which will allow it to be compared in detail with numerical relativity simulations: by the time of merger all parameters are measured. Finally, the ringdown signal is fit to the quasi-normal mode (QNM) frequencies of Kerr holes, determining the mass and spin of the final black hole.

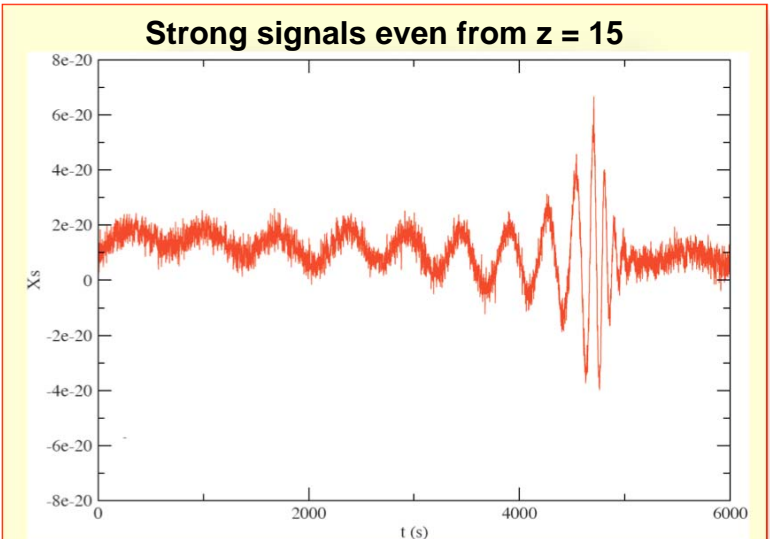

Figure 1: The GW signal for the final few orbits, plunge, merger and ringdown of a system consisting of two $10^5 M_\odot$ nonspinning MBHs at redshift $z = 15$, seen face-on. The signal contains simulated LISA noise. Time is measured in seconds. Note that even at $z = 15$, the waveform stands up well above the noise and is visible in fine detail (Baker *et al* 2007).

Any serious lack of fit in any of these stages would constitute a difficulty for GR. Although there are many parameters to fit for the inspiral, the phase has typically $10^4$—$10^5$ cycles, lasting months or years, so there is far more data than parameters. Departures from GR even at high PN order that change the number of cycles by just ±1 would be detectable. One way the fit could fail is if the graviton has a small mass, causing dispersion among the frequency components of the inspiral signal (Berti *et al.* 2006). LISA would be sensitive to masses $10^4$ times smaller than the present bound of around $10^{-22}$ eV.

**BH mergers: brighter than the entire universe**

The SNR for the BH merger phase is very high, even though there are few cycles, because black-hole mergers are energetically the brightest events in the universe. Each event radiates GWs at about $10^{23} L_\odot \sim 10^{56}$ erg/s, independent of the masses of the holes. This is more than the integrated luminosity of the rest of the entire universe!

Numerical simulations of merger are already very accurate: typical accumulated phase errors are smaller than one degree over the full merger. By the time LISA flies it should be straightforward to use the enormous merger SNR of order $10^6$ or more (see box) to find even small deviations from GR theory. If the inspiralling objects are, say, boson stars rather than black holes, then the merger and ringdown radiation would be completely different from that expected from MBHs.

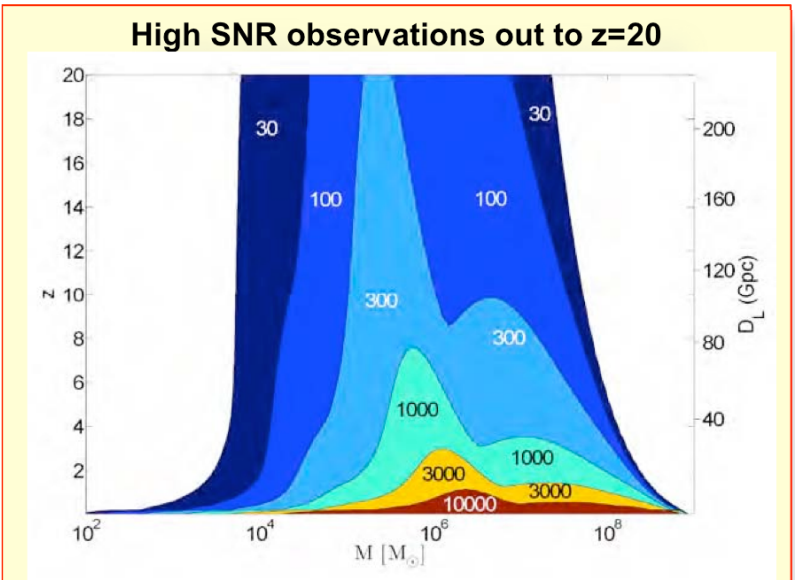

Figure 2: Contour plot of amplitude SNR (square it to get energy SNR) for the entire inspiral and merger signal, for equal-mass nonspinning binaries, as a function their total mass and redshift (Baker *et al.* 2006). For tests of GR the closest/strongest (red) sources give the cleanest results.

The information from the ringdown phase is a crucial test. For each detectable quasi-normal modes, LISA measures both the frequency and the damping rate, and these in turn determine both the mass and the spin of the hole. Thus, all the modes must be consistent if the remnant is a Kerr black hole. They also test whether the Hawking area theorem holds (is the area of the final Kerr hole larger than the sum of the areas of the inspiralling holes?). If the final object is not Kerr, then the ringdown modes are clues to its nature. And if there are no modes at all, the remnant could be a naked singularity!

It is important that alternative theories of gravity that have massless scalar fields in addition to the metric generally have black holes identical to Kerr: the scalar field is radiated away when the hole forms, according to the no-hair theorem. So any evidence for extra radiation during the long inspiral phase would indicate that the massive objects were not black holes at all.

## Extreme mass ratio inspirals: Precision probes of Kerr spacetime

Observational evidence for the existence of MBHs at the centers of galaxies is currently based on modeling the gravitational potentials of these objects using the motions of stars and gas, and comparing the results with those expected if the central object were a black hole. The best case today comes from stellar motions near the center of our galaxy, which reveal the presence of a compact dark object of mass $M \sim 4 \times 10^6 M_\odot$; the orbits show the central mass to be pointlike down to a scale of

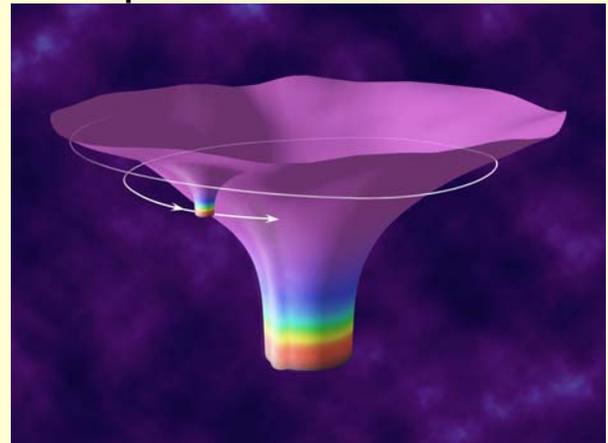

Figure 3: Embedding diagram of an EMRI, with the smaller black hole orbiting in the spacetime of the larger black hole. The colors depict the slowing of time (the "lapse" function) as one nears the horizons and the shape depicts the geometry of space in the orbital plane.

~100 AU. Using milliHertz GWs emitted by stellar-mass black holes and other compact stellar probes that are captured by massive central objects, LISA will map the spacetime around these central objects down to length scales ~$10^4$ times smaller – the size of the horizon.

Such a capture inspiral with one body much less massive than the other is referred to as an extreme-mass-ratio inspiral (EMRI). The emitted GW frequency is determined by the mass of the central black hole, so LISA could study central objects with masses of $10^5 – 10^7\ M_\odot$ through capture events with mass ratios $10^{-7} < m_2/m_1 < 10^{-2}$. Stellar-mass BHs, shown in Figure 3, are expected to dominate the rate.

Signals from EMRIs are generally small enough to require matched filtering in order to extract them from the data stream. LISA will observe each EMRI for a timescale of years, or equivalently for ~$10^5$ cycles, which raises the matched filtering power SNR by the same factor of $10^5$ above the instantaneous signal-to-noise ratio. The parameter space for EMRI filters is large enough that brute-force filtering will challenge even the supercomputers of ten years from now. The LISA community is therefore developing more efficient near-optimal algorithms, which are being tested in the Mock LISA Data Challenges (Babak, et al, 2008). The best current estimate, extrapolating from an estimated EMRI rate in the Galaxy of $2.5 \times 10^{-7}$/yr (Hopman 2006), is that LISA will have an EMRI detection rate of ~ 50 - 100/yr, with the strongest sources having power SNR exceeding $10^4$ (Gair *et al.* 2004).

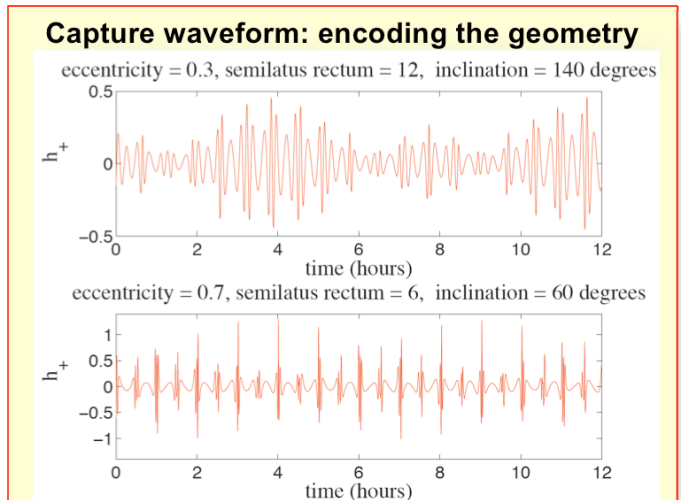

Figure 4: Segments of generic noise-free EMRI waveforms (Drasco & Hughes 2006) produced by a test mass orbiting a $10^6 M_\odot$ black hole that is spinning at 90% of the maximal rate allowed by general relativity. The top panel assumes a slightly eccentric and inclined retrograde orbit modestly far from the horizon. The bottom panel assumes a highly eccentric and inclined prograde orbit much closer to the horizon. The amplitude modulation visible in the top panel is mostly due to Lense-Thirring precession of the orbital plane. The bottom panel's more eccentric orbit produces sharp spikes at each pericenter passage.

EMRI signals will be a powerful new testbed for GR for two reasons. The first is that all $10^5$ cycles of the radiation are emitted from near the MBH horizon. The reason is that the radiation-reaction time-scale varies inversely with the mass ratio of the captured object to the MBH. Therefore, a 10 $M_\odot$ BH capture by a $10^6\ M_\odot$ MBH moves inward in its last $10^5$ orbits the same distance that two equal-mass $10^6\ M_\odot$ MBHs move in their very last orbit. EMRIs probe the strong-field region all the time they are being observed. The second reason that EMRIs are important for tests is the complexity of their orbits. The strong influence of the Kerr spin prevents planar orbits: a generic orbit spirals over much of a sphere surrounding the center. The resulting emitted waveform reflects this complexity (Figure 4), which is a treasure-store of information about the geometry.

EMRIs are expected to be very clean astrophysical systems (except perhaps in the few percent of galaxies containing active galactic nuclei, where interactions with the accretion disk could possibly affect the dynamics). The Kerr metric is described by just two parameters, mass and spin; the filter that extracts the signal will give best-fit values of these numbers and of data about the captured black hole: its mass, its spin, and the initial orbital parameters. But with power SNRs of $10^3$-$10^4$ after filtering, small deviations of, say, the geometry's higher multipole moments from those of Kerr would be detectable.

If the central object is not a black hole, but rather a boson star or something similar, then the inspiraling object will continue to emit long after it would shut off in Kerr (Kesden *et al.* 2005). This would be a clean and blindingly simple falsification of the central black hole paradigm.

## Perspective

Gravitational wave detection will not only open a new information channel on known systems; even more importantly, it will begin to explore a completely unseen part of the universe. Exploration like this has often transformed physics: new discoveries unexpectedly overturn accepted models, disprove accepted theories. The first GW detections will be made by the ground-based instruments, LIGO and VIRGO, at frequencies above 100 Hz. They are likely to see a number of binary merger events, many involving stellar-mass black holes. They will test strong-field GR to at least the 10% level, perhaps better. But when LISA opens up the milliHertz frequency band, where sources are intrinsically stronger, it will become the first detector with high enough SNR to be able to make true high-precision GW measurements. As with the perihelion shift of Mercury, the Lamb shift, the monitoring of the Binary Pulsar: new physics is sometimes hidden in the $n^{th}$ decimal place of a measurement on a system that is clean enough that explanations based on old physics can be excluded. LISA's high-precision observations of clean systems governed by nonlinear dynamical gravity give it enormous discovery power. It will expand the scope of astronomy and physics significantly and may well lead to new paradigms in the understanding of our universe.